\documentclass{article}

\usepackage{url,graphicx,subcaption,comment}
\usepackage[]{fullpage}
\usepackage{multicol,float}

\newcommand{\eg}{\emph{e.g.}~}
\newcommand{\ie}{\emph{i.e.}~}
\newcommand{\etal}{\emph{et al.}~}
\newcommand{\wrt}{\emph{wrt.}~}

\usepackage{titling}
\begin{document}

\title{A survey to measure cognitive biases\\ influencing mobility choices}
\author{Carole Adam}
\maketitle

\abstract{
In this paper, we describe a survey about the perceptions of 4 mobility modes (car, bus, bicycle, walking) and the preferences of users for 6 modal choice factors. This survey has gathered 650 answers in 2023, that are published as open data. In this study, we analyse these results to highlight the influence of 3 cognitive biases on mobility decisions: halo bias, choice-supportive bias, and reactance. These cognitive biases are proposed as plausible explanations of the observed behaviour, where the population tends to stick to individual cars despite urban policies aiming at favouring soft mobility. This model can serve as the basis for a simulator of mobility decisions in a virtual town, and the gathered data can be used to initialise this population with realistic attributes. Work is ongoing to design a simulation-based serious game where the player takes the role of an urban manager faced with planning choices to make their city more sustainable. 
}

\section{Introduction}

Mobility is a central issue in the transition to a more sustainable lifestyle. The average daily distance traveled by the French population has increased considerably, from 5 km on average in the 1950s to 45 km on average in 2011 \cite{viard2011eloge}, as has the number of personal cars (11,860 million cars in 1970 \cite{barre1997some} compared to 38,3 million in 2021 \cite{ccfa2019,insee}). For example in Toulouse, cars concentrate 74\% of the distances traveled by the inhabitants and contribute up to 88\% to GHG emissions \cite{toulouse}. The evolution of mobility is therefore an essential question, both for the global climate crisis and for public health: negative impact of a sedentary lifestyle \cite{biswas2015sedentary}, road accidents, air and sound pollution \cite{eea2016}. Indeed, 40000 deaths per year are attributable to exposure to fine particles (PM2.5) and 7000 deaths per year attributable to exposure to nitrogen dioxide (NO2), \ie 7\% and 1\% of the total annual mortality \cite{spf2021}; the 2-month lockdown of spring 2020 in France saved 2300 deaths by reducing exposure to particles, and 1200 more deaths by reducing exposure to nitrogen dioxide \cite{spf2021}.

This shows that public policies and individual behaviour changes (modal shift towards cycling, more extensive teleworking) can have an impact on public health. For instance during the COVID-19 pandemics many temporary cycling lanes were set up \cite{rerat2022cycling}, although many have since been returned to cars \cite{barbarossa2020post}. Aside from such emergencies, public policies take time to set up, and they are not always well accepted. Indeed, despite feeling more and more concerned about climate change, citizens are often reluctant to constraining public policies that could slow it down, as shown by recent strikes against petrol taxes or new road tolls. As a result, mobility evolves very slowly, for instance in France a large proportion of commuting is still done by car, even for very short journeys \cite{brutel2021voiture}. Many explanations of this inertia were proposed: constrained use of the car (\eg to transport children or tools); lack of alternatives (limited public transports); social inequities (car-dependency in rural areas, cycling facilities and public transport concentrated in town centres, cost of electric or newer cars...); difficulty of changing habits \cite{brette2014reconsidering,lanzini2017shedding}; individualism \cite{epprecht2014anticipating}; or cognitive biases influence on human reasoning \cite{innocenti2013car,bcg2020,isaga-bias}. Mobility choices are therefore explained by a combination of the context (infrastructures and services provided where the user lives and works), individual characteristics (age, gender, fitness, handicap), and psychological and sociological factors (preferences, emotions, biases, social pressure). Urban policies can modify the infrastructure (develop public transport or cycling lanes, for example), adapt their offer to various user profiles (accessibility, gender, financial help), or communicate to try and modify individual preferences and choices (for example campaigns in favour of active mobility for health reasons, advocating for sustainable modes for ecology reasons, promoting electric cars as less expensive on the long term, promoting road safety, or a mix of these arguments). But an essential aspect yet neglected in modality policies is the consideration of cognitive biases influencing individual choices \cite{bcg2020,innocenti2013car}.

In order to highlight some of the biases involved in mobility decisions, we conducted a survey about people's subjective perceptions of mobility. We asked people about their decision factors, \ie what is important to them when choosing a mobility mode (ecology, price, etc). We also asked them how they perceive the different mobility modes available (\eg how safe is it to cycle, how expensive is the car, etc). Our survey collected 650 answers during March-July 2023. This paper presents a first analysis of the results, including a study of the cognitive biases that could explain apparently irrational choices. We also show some lessons that can be learned for mobility policies.

The paper is structured as follows. In Section~\ref{sec:soa} we analyse the literature about mobility in sociology, psychology, and computer modeling. Section~\ref{sec:survey} describes our survey methodology, while Section~\ref{sec:results} presents a statistical analysis of the answers. 
Section~\ref{sec:biases} then focuses on the identification of three specific cognitive biases to explain the mobility choices observed. Finally, Section~\ref{sec:discuss} discusses our results, how they can be used to inform mobility policies, and their limitations, before Section~\ref{sec:cci} concludes the paper.

\section{Background} \label{sec:soa}

\subsection{Sociology of mobility}

\paragraph{User profiles.}
The Mobil'Air study \cite{chalabaevc} presents 4 car user profiles: open to all modes; attached to an individual mode, ensuring independence; constrained; and convinced, loving to drive, for all their journeys. This study provides percentages of the population in each category, but does not consider other modes of mobility. The study also notes the importance of constraints depending on the reason for travel (transporting children for example), and the strength of habits or routines.

Rocci \cite{rocci2007automobilite} proposes 6 user profiles: passionate car drivers adhering to the car; passionate car drivers in opposition to another mode; rational multimodal users, sticking to the car but sometimes using public transport; multimodal users in opposition to the car but sometimes forced to take it; alternative mode users passionate about their mode; and alternative mode users in opposition to their mode. This classification is more detailed and considers other modes than the car, but does not provide population distribution statistics. We find several similarities with the previous survey, such as constrained or on the contrary passionate use of the car. In particular, Rocci shows that driving pleasure allows to accept or forgot the associated constraints (delays, parking difficulty, cost...), while a mode considered as only utilitarian will be switched for another one as soon as it does not fill its function anymore. This is a strong differences between modes whose use can be passionate (car or bicycle) vs public transport. Finally, Rocci also shows the inter-individual differences in perception of modes: for example, convinced car drivers tend to underestimate the price of the car, and overestimate that of public transport.

Another important point in mobility profiles is the role of gender \cite{pelgrims2023gendered}. Women often feel more exposed in their use of public modes, and feel more concerned with safety.

\paragraph{Decision factors.}
In addition to environmental constraints, the modal decision also depends on each person's 'mobility capital' \cite{rocci2007automobilite} (owning a bicycle, having a driving license, being fit to cycle or walk, living close to a train or bus stop...). Beyond these constraints, everyone will evaluate different aspects, such as price, safety, or travel time. The choice must also minimise the mental load (number of connections, for instance). Public transport has an image associated with stress, risk of aggression and dependence \cite{kaufmann2010si}, unlike the car which conveys an image of autonomy and social success, reinforced by advertisements. 

In our survey described in this paper we investigate the priorities and perceptions of 6 decision criteria, as extracted in previous work \cite{ajce22a}: cost, time, practicality, safety, comfort, and ecology. Other dimensions that could be considered in the future include the health impact, often promoted for 'active' mobility modes. \cite{weyer2023modeling} conducted a survey in Germany with 10782 people, and retained very similar dimensions: mobility modes were rated as being cheap, fast, environmentally friendly, comfortable, safe and reliable. They have no practicality dimension (which represents the cognitive load associated with the mode) but a reliability criteria, representing the probability that the trip fails. They then ran a clustering analysis on the values of these dimensions in their sample, to deduce 5 distinct user profiles, such as comfort-oriented or price-sensitive.

\paragraph{Habits.}
Mobility choices are not only a result of these rational dimensions. Habits are at the heart of mobility decisions \cite{brette2014reconsidering}: individuals tend to reproduce habitual decisions when they are in the same context. This process can save decision time, but also lead to decisions that are not adapted to changes in the environment, if they are not reconsidered. Habits can also modify perceptions (time, cost, etc.): thus a user accustomed to taking the car can overestimate travel time by public transport and under-estimate travel time by car \cite{rocci2007automobilite,betsch2001effects}.

Life cycle changes, during which all habits are disrupted (job change, moving, birth, etc.), are favorable moments for breaking old habits and creating new ones \cite{rothman2015hale}. Work has thus experimented with the distribution of free public transport tickets to newly arrived residents, to encourage them to abandon the car \cite{bamberg2003does}. The COVID-19 pandemic has also shown an unusual reset of habits and encouraged bicycle travel, at least temporarily \cite{barbarossa2020post}.

\subsection{Cognitive biases}

\paragraph{Definition.}
Cognitive biases are heuristics used by our cognitive system to facilitate decision-making in situations of uncertainty or danger \cite{tversky1974judgment}. They allow rapid reasoning during stressful or complex situations despite the incompleteness or uncertainty of the information necessary for rational decision-making. They are thus useful to make faster decisions despite the lack of information, and are therefore essential to our proper functioning. But as any heuristics, they can lead to mistakes, seemingly irrational decisions, with sometimes serious consequences. 

For instance \cite{crash1982} analyse a plane crash as resulting from the pilot's self-deception bias that prevented him from reacting to alerts from his copilot. \cite{fouillard2021catching} propose a logical model of biases allowing to propose explanations of erroneous decisions leading to accidents. \cite{murata2015influence} show that it is necessary to recognize and eliminate biases to avoid various accidents or collisions. \cite{doherty2020believing} highlight the importance of educating medical personnel about biases that can affect their decisions and diagnoses. \cite{luz2020heuristics} list various biases that affect patient decisions about vaccination. Finally \cite{mazutis2017sleepwalking} show how biases can explain companies' lack of adaptation to climate change. This shows that cognitive biases must be taken into account when trying to understand or influence human decisions.

\paragraph{Biases and mobility.}
Innocenti \etal \cite{innocenti2013car} show that people tend to 'stick' to the car, even if more expensive than metro or bus, and explain this irrationality by the influence of cognitive biases. They conclude that mobility policies need to try to modify the perception of the different modes of mobility.

Another study \cite{bcg2020} looks at the reasons why drivers are generally reluctant to switch from their personal car to new modes of mobility, even those proven to be more efficient. Although this study focuses on carpooling or free-floating bikes and scooters, their findings are interesting. They find that ''seemingly irrational'' mobility decisions ''that work against the new mobility'' are influenced by various emotions, social norms and cognitive biases, and are not taken into account by mobility operators. The \textbf{halo bias} pushes motorists to amplify the benefits of driving (autonomy...) and ignore its disadvantages (delays in traffic jams...). The \textbf{ambiguity bias} pushes them to prefer known risks to unknown risks, to maintain an illusion of control, and thus to avoid the uncontrollable risks posed by certain modes (delays or breakdowns of public transport). The \textbf{anchoring bias} implies that a negative first impression regarding a new mode of mobility will be retained, preventing future reuse. The \textbf{status quo bias} induces a preference to keep things as they are, in order to save cognitive load, similarly to habits \cite{brette2014reconsidering}. The study also lists emotional and social factors which explain this attachment to the car: pride associated with owning a car, aggressiveness towards users of other modes \cite{delbosc2019dehumanization} and fear of becoming the target in the event of a modal switch; or fear of crime, particularly among women.

They also list emotional and social factors that explain car stickiness, such as car pride bias (owning a nice car is valued); fear of the unknown (new modes are considered more dangerous by non-users than by users); negative social emotions towards outgroups or disruptors (users of other, in particular new, modes), who can be the target of aggressive behaviour \cite{delbosc2019dehumanization}; and symmetrical fear of being themselves targeted by such aggressions from their current social group if they were to switching to new disrupting modes; and fear of crimes, where women in particular are found to be more concerned with threats to their safety in public spaces, such as using the public transports or sharing rides.

Other biases that could be involved include \textbf{reactance} \cite{brehm1966theory}, defined as the tendency to react to persuasion attempts felt as coercive by asserting one's free will and strengthening one's non-compliant position as a result. Such messages could thus have an effect contrary to that intended, by agonising receivers. For instance \cite{sakai2021psychological} show that restrictions of mobility during the pandemics led people to a higher desire of going out; while \cite{luo2023restriction} show that the loss a freedom felt during lockdowns leads to preferences for advertisements for tourism products. \cite{isaga-bias} proposed a computer model of the influence of this bias on mobility choices.

\textbf{Choice-supportive bias}, or \emph{a posteriori} rationalisation or post-purchase rationalisation \cite{mather2000misremembrance} consists in rationalising a choice after it was made, by attributing positive features to the option selected, and more negative features to the rejected options. It is similar to the halo bias discussed above. This modified interpretation aims at reducing one's cognitive dissonance and regret; for instance someone who drives to work can insist on the autonomy provided by the car, or on the dangers associated with cycling.

\section{Survey methodology} \label{sec:survey}

\enlargethispage{30pt}
\paragraph{Objective.}
The goal of the survey is to understand the priorities and perceptions of people when evaluating mobility options. We also want to evaluate how rational their decision is, \ie to compare their declared usual mobility mode with predictions made by a rational decision algorithm based on their personal priorities and perceptions. When their choice is deemed irrational, we will try to reveal the impact of various cognitive biases in this difference. We will also investigate gender differences in these perceptions and choices.

\paragraph{Questions.}
Our questionnaire consists of three main parts. The first part concerns the responders' profile and their mobility habits. In the second part, the participants are asked to provide their priority ratings over the 6 decision criteria identified above (ecology, comfort, financial accessibility, practicality, safety and speed), \ie to state how important each criterion is in their daily mobility choice. The third part concerns the participants' perceptions of the value of the mobility modes considered (bicycle, car, public transport and walking) over these criteria, \ie to mark how well they think that each mode satisfies each criterion. All ratings (priorities and values) are made on a Likert scale from 0 to 10. The questionnaire was administered in French. Below is a translation of its main elements:\begin{footnotesize}
\texttt{
\begin{itemize}
    \item Part 1: responder profile 
    \begin{enumerate}
        \item What is your gender (Multiple choice: woman, man, other, do not wish to answer)
        \item Which mobility mode do you use most for your daily trips? (Choice: bike, car, public transport, walk)
        \item What is the distance between your place of residence and your main place of activity, in km? 
        \item How many times per week do you realise the roundtrip between your residence and your place of activity? 
        \item Are any mobility modes inaccessible to you (personal constraints, lack of infrastructures, for instance no public transport available)? (Multiple choice: bicycle, car, public transport, walking, none)
        \item Give precisions if you wish. (free text)
    \end{enumerate}
    \item Part 2: importance of choice criteria: for each criteria, we ask you to evaluate how important it is for you when choosing your daily mobility mode. Answers on a Likert scale from 0 to 10 for the following 6 criteria:
    \begin{itemize}
        \item Ecology: this mode has a low carbon footprint
        \item Comfort: this mode is pleasant to use
        \item Financial accessibility: this mode is cheap
        \item Practicality: this mode is flexible, does not impose strong constraints 
        \item Safety: this mode allows me to move without risks (accidents, injury, aggression)
        \item Speed: this mode allows me to reach my destination in low time
    \end{itemize}
    \item Part 3: evaluation of responder's perceptions of mobility means. For each of the 4 modes:
    \begin{enumerate}
        \item How would you rate the (ecology, comfort, financial accessibility, praticity, safety, speed) of this mode as a daily commuting mode? 6 questions, answers on a Likert scale from 0 to 10.
        \item Do you want to add anything on your perception of this mode? (free text answer)
    \end{enumerate}
\end{itemize}
}
\end{footnotesize}

\paragraph{Privacy.}
The questions were verified by a privacy researcher and the university DPO, to ensure compliance with personal data regulations. No question allows to identify the responders, in particular we do not collect names, emails, addresses of work or residence, or age. The only sensitive data collected is the gender, which was needed for the purpose of our study.

\paragraph{Targeting.}
The survey was administered through an online form. This form was circulated on various mailing lists addressed to researchers, university students, or personal contacts, in France. Our sample is therefore not fully representative of the global population, but biased towards people working or studying in universities. Although no question allows us to know the geographical origin of an answer, we believe that the sample is also geographically biased, with more answers from Grenoble, where the author works.

\paragraph{Raw data.}
The anonymous answers obtained from this survey are publicly available as open data, to allow reproduction of this research \cite{ELLXJF_2024}. To ensure protection of responders' privacy, the free comments are not published, because they often contain sensitive data (geographical origin, handicap, etc).

\section{Survey results analysis} \label{sec:results}

\subsection{Statistical results}

\paragraph{Mobility distribution}

\begin{figure}[ht]
    \centering
    \begin{subfigure}{0.45\textwidth}
        \centering
        \includegraphics[scale=0.4]{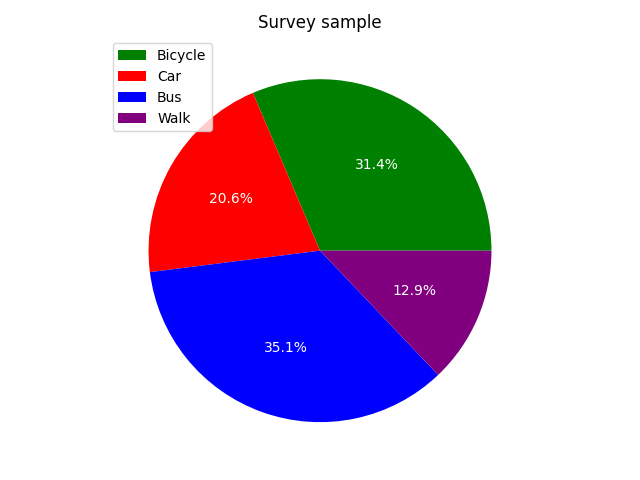}
        \caption{Our sample: \\car 20.6\%, bike 31.4\%, bus 35.1\%, walk 12.9\%}
        \label{fig:pie1}
    \end{subfigure}
    \begin{subfigure}{0.45\textwidth}
        \centering
        \includegraphics[scale=0.4]{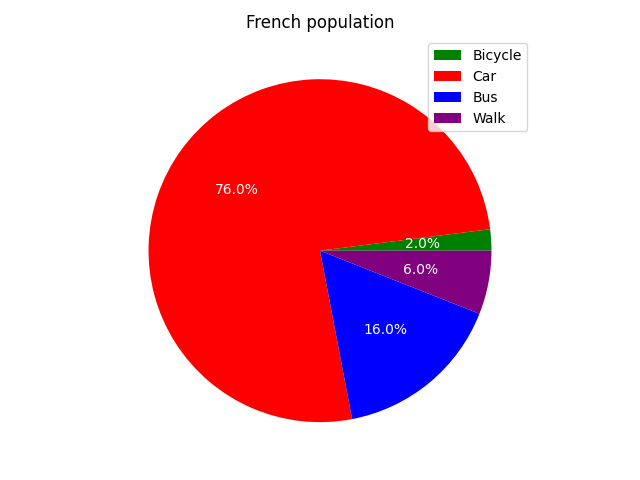}
        \caption{France 2023, data from \cite{perona2023deplacements}:\\ car 76\%, bike 2\%, bus 6\%, walk 16\%}\label{fig:pie2}
    \end{subfigure}
    \caption{Distribution of mobility: comparing our sample with the French population}
\end{figure}

The respondents did provide their usual mobility mode. Figure~\ref{fig:pie1} shows the distribution of answers over the 4 options in our sample. Comparatively, Figure~\ref{fig:pie2} shows the modal distribution in the general French population, based on \cite{perona2023deplacements}. We can notice that our sample is not representative. This is due to our targeting: first, geographically, the authors live in areas where the bicycle is over-represented compared to the national average\footnote{INSEE statistics for Grenoble: \url{https://www.insee.fr/fr/statistiques/2557426}}; second, socially, our survey was mainly answered by students and university staff, who are not representative of the global population. However, we have obtained a large representation of each mobility mode, which will allow us to draw significant quantitative conclusions. 

\paragraph{Accessibility.}

Figure~\ref{fig:access} below shows some statistics about the accessibility of modes. Figure~\ref{fig:access-bar} splits the accessibility results per mode, showing the count of users who cannot access each specific mode; please note that the sum is not equal to the total number of answers since some users report no inaccessible mode, or on the contrary report several ones. Figure~\ref{fig:access-pie} details the percentage of users by number of modes that are inaccessible to them. All users have at least 1 mode accessible, but many answers report at least one mode being inaccessible.

\begin{figure}[ht]
    \centering
    \begin{subfigure}{0.45\textwidth}
        \centering \vspace*{-10pt}
        \includegraphics[scale=0.4]{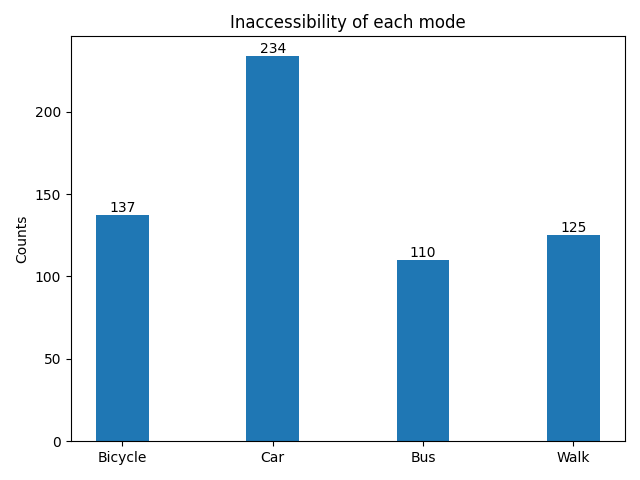}
        \caption{Number of users without access to each mode}
        \label{fig:access-bar}
    \end{subfigure}
    \begin{subfigure}{0.45\textwidth}
        \includegraphics[scale=0.4]{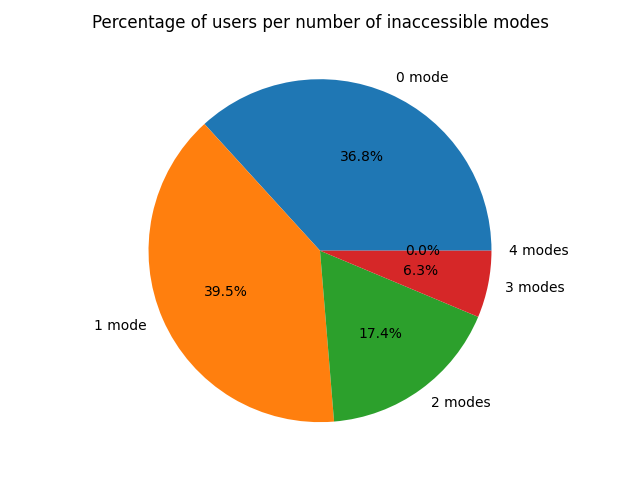}
        \caption{Number of inaccessible modes}
        \label{fig:access-pie}
    \end{subfigure}
    \caption{Accessibility of modes}\label{fig:access}
\end{figure}

\newpage
\subsection{Mean values}

\paragraph{Priorities.}

Table~\ref{tab:priorites} reports the average priorities of our 6 criteria in the population of respondents (n = 650), and in the sub-populations of users of each mode. We can observe differences between users of different modes, with some criteria being more salient or in the contrary more insignificant for users of a given mode. For instance cyclists have a much higher priority for ecology, and drivers a much lower one; walkers tend to neglect time, and focus on price, along with bus users. These differences in priority profiles will be discussed later. These average values can be used to initialise a synthetic population for an agent-based simulation, thus endowing agents with realistic priority profiles \cite{conrad2024identifying,weyer2023modeling}. Future work would be needed to do a clustering analysis and deduce types of users; for the moment we only regroup users of the same mode. \\\strut\\

\begin{table}[ht] 
\centering 
\begin{tabular}{|c|c|c|c|c|c|} 
\hline 
 & all (650) & Bicycle (204) & Car (134) & Bus (228) & Walking (84) \\ 
\hline\hline 
Ecology & 7.08 & 8.3 & 5.65 & 6.76 & 7.27\\ 
 \hline 
Comfort & 7.1 & 7.31 & 7.19 & 6.75 & 7.35\\ 
 \hline 
Finance & 6.97 & 7.08 & 5.63 & 7.44 & 7.58\\ 
 \hline 
Practicality & 8.27 & 8.54 & 8.57 & 7.81 & 8.42\\ 
 \hline 
Time & 7.47 & 7.68 & 7.79 & 7.37 & 6.7\\ 
 \hline 
Safety & 6.2 & 5.37 & 6.72 & 6.46 & 6.67\\ 
 \hline 
\end{tabular} 
\caption{Average priorities of criteria, over all answers, and over users of each mode} 
\label{tab:priorites} 
\end{table}

\paragraph{Evaluations.} Beyond priority profiles, individuals also differ in their evaluations of how well a given mobility mode fits a given criterion. Table~\ref{tab:evals} reports the average values of Bicycle, Car, Bus and Walking over the 6 criteria. The values are averaged over the global population, over the sub-population of users of this mode, and over the sub-population of non-users of this mode (users of any one of the other 3 modes). This allows to notice a difference in evaluations between users and non-users, where users of a given mode systematically evaluate it higher than the rest of the population on any criteria. This is particularly obvious for the comfort and practicality, for the cost of driving, or for the speed of any given mode. These differences will be discussed in more details in the following. 

\begin{table}[ht]
\small
    \centering 
    \begin{subtable}{0.3\textwidth}
        \begin{tabular}[t]{|c|c|c|c|c|}
        \hline
        Bicycle & All & Users & Non-u \\ 
        \hline
        Ecology & 9.21 & 9.56 & 9.05 \\
            \hline
        Comfort & 6.03 & 7.39 & 5.4 \\
            \hline
        Finance & 7.74 & 8.54 & 7.37 \\
            \hline
        Practicality & 6.63 & 8.23 & 5.9 \\
            \hline
        Time & 6.6 & 7.98 & 5.96 \\
            \hline
        Safety & 4.62 & 5.38 & 4.28 \\
            \hline
        \end{tabular}
        \caption{Bicycle (n=204) }
    \end{subtable} \qquad \qquad \qquad 
    \begin{subtable}{0.3\textwidth}
        \begin{tabular}[t]{|c|c|c|c|c|}
        \hline
        Car & All & Users & Non-u \\ 
        \hline 
        Ecology & 1.81 & 2.52 & 1.63 \\
            \hline
        Comfort & 7.69 & 8.51 & 7.47 \\
            \hline
        Finance & 2.68 & 3.84 & 2.38 \\
            \hline
        Practicality & 6.32 & 8.32 & 5.81 \\
            \hline
        Time & 6.76 & 8.21 & 6.38 \\
            \hline
        Safety & 7.29 & 7.69 & 7.19 \\
            \hline
        \end{tabular}
        \caption{Car (n=134)}
    \end{subtable}
\\ \strut \\
\strut \centering
    \begin{subtable}{0.3\textwidth}
        \begin{tabular}[t]{|c|c|c|c|c|}
        \hline
        Bus & All & Users & Non-u \\ 
        \hline 
        Ecology & 7.43 & 7.77 & 7.25 \\
        \hline
        Comfort & 5.83 & 6.46 & 5.49 \\
        \hline
        Finance & 6.87 & 7.25 & 6.66 \\
        \hline
        Practicality & 5.78 & 7.2 & 5.0 \\
        \hline
        Time & 5.57 & 6.81 & 4.91 \\
        \hline 
        Safety & 7.46 & 7.37 & 7.5 \\
        \hline
    \end{tabular}
    \caption{Bus (n=228)}
    \end{subtable}  \qquad \qquad \qquad
    \begin{subtable}{0.3\textwidth}
        \begin{tabular}[t]{|c|c|c|c|c|}
        \hline
        Walk & All & Users & Non-u \\ 
        \hline 
        Ecology & 9.81 & 9.74 & 9.83 \\
        \hline
        Comfort & 6.7 & 8.12 & 6.49 \\
        \hline
        Finance & 9.75 & 9.79 & 9.74 \\
        \hline
        Practicality & 5.99 & 8.01 & 5.69 \\
        \hline
        Time & 2.98 & 4.96 & 2.69 \\
        \hline
        Safety & 6.77 & 7.12 & 6.72 \\
        \hline
    \end{tabular}
    \caption{Walk (n=84)}
    \end{subtable}  
 
    \caption{Evaluations of 4 modes on 6 criteria}
    \label{tab:evals}
\end{table}
\normalsize

\newpage
\subsection{Analysing decisions} \label{sec:ratio}

\paragraph{Mobility scores.}

The scores of the 4 mobility modes are not directly given by responders, but computed as a weighed average of their own evaluations of modes on criteria, weighted by their personal priorities for these criteria. Concretely, an individual $i$ has personal priorities for each criterion $c$, denoted $prio_i(c)$ (how important this criterion is), as well as personal values of mode $m$ over each criterion $c$, denoted $val(m,c)$ (how well they believe that mode $m$ satisfies criterion $c$); we use these to deduce the personal score of mode $m$ for individual $i$ with the following formula:
$$score_i(m) = \sum_{c \in crits} val(m,c) * prio_i(c)$$

Table~\ref{tab:stats-modes} summarises statistics about the scores obtained for the 4 mobility modes in our sample. The table provides the average score over all answers, standard deviation, median score, as well as the average scores over users of the mode vs non-users. By users, we mean people who declared that mode as their usual commuting mode, while non-users are those who declared another usual mode. We can see that the evaluations for some modes are very variable, with standard deviations up to 1.75 points for marks given on a scale of 10. Comparing marks between users and non-users also shows wide differences. In the following we investigate these differences in more details.

\begin{table}[ht]
    \centering
    \begin{tabular}{|c|c|c|c|c|c|}
        \hline
        Mode & Mean mark & Stdev & Median & Users & Non-users \\
        \hline
         Bicycle & 6.85 & 1.66 & 7.06 & 8.11 & 6.27 \\
         \hline
         Car & 5.41 & 1.75 & 5.47 & 6.93 & 5.01 \\ 
         \hline
         Bus &  6.43 & 1.47 & 6.62 & 7.21 & 6.01 \\ 
         \hline
         Walk & 6.90 & 1.52 & 7.07 & 8.00 & 6.73 \\ 
         \hline
    \end{tabular}
    \caption{Marks given to the 4 modes of mobility}
    \label{tab:stats-modes}
\end{table}

\paragraph{Constrained choice.}

In our survey, the responders provided their perception of availability for the 4 modes. We used these answers and checked for each responder if the mode that receives the best score according to the formula above is considered to be unavailable to them. In the case when the best option (\wrt their personal priorities and evaluations) is not available (\wrt to their own perceptions), we consider that this responder's choice is constrained. It is important to note that unavailability can be subjective: different users will have different distance thresholds to consider that their work is too far to walk or cycle, for instance, or different thresholds regarding how complicated a public transport trip can be before it is considered unfeasible. This unavailability can also differ between users in the same context, for instance public transport can be available in the area, but not adapted to some kinds of handicap.

\paragraph{Rational choice.}
Finally, we define the rational modal choice, in line with \cite{jacquier2021choice}, as the mobility mode, among those available to the user, that moximises the score computed above. For each user, we compared the mode receiving the best score (excluding unavailable modes) to the mode that they declared as their usual commuting mode. If they match, the choice is deemed rational: this responder did choose the available mode that receives the best score; if they differ, meaning that the user does not use a mode that is available and receives a better score, then the choice is deemed irrational. Individuals with different priorities will have different rational choices even though they are in the same environment. It is important to notice that the choice is deemed 'rational' on the basis of subjective evaluations that could actually be biased and help rationalise a choice that is not so rational. Below we show how these subjective perceptions result in differences between genders and between habitual users of different modes. Later, Section~\ref{sec:biases} will investigate if the mobility choices reported by the responders are rational in the sense of that model, and if not, why.

\subsection{Gender differences} \label{sec:gender}

\paragraph{Priorities.}
Figure~\ref{fig:gender-prio} details the priority profiles for male vs female responders. The sum of answers from men and women is not equal to 650 users, because a number of answers have selected another gender, or preferred not to answer; however, there were not enough results for these options to draw significant statistics, so we omitted them here. We can see that priorities are very similar on most criteria, except for 2 significant differences: females give more priority to comfort and safety than males. This could be partially explained by the fact that women more often transport their children \cite{nobis2005gender}. 

\begin{multicols}{2}
\begin{minipage}{0.45\textwidth}
\vspace*{-25pt}
\begin{figure}[H]
    \centering
    \includegraphics[scale=0.37]{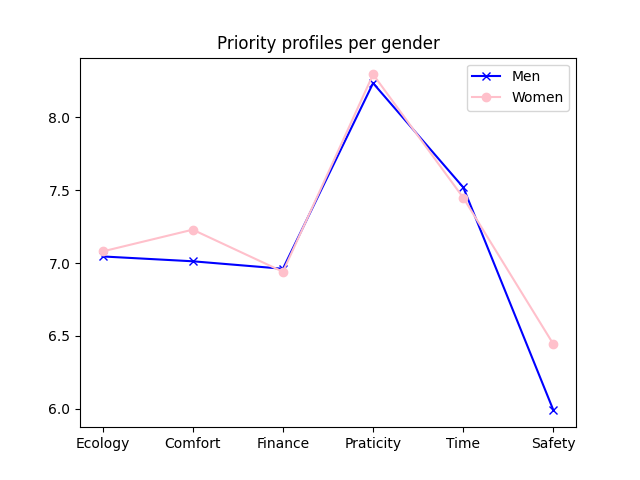}
    \caption{Gendered priority profiles}
    \label{fig:gender-prio}
\end{figure}
\end{minipage}

\begin{minipage}{0.45\textwidth}
\begin{table}[H] 
\centering 
\small
\begin{tabular}{|c|c|c|c|} 
\hline 
 & All & Men & Women\\ 
\hline\hline 
Number & 650 & 331 & 301\\
\hline
Ecology & 7.08 & 7.05 & 7.08\\
\hline
Comfort & 7.1 & 7.01 & 7.23\\
\hline
Finance & 6.97 & 6.96 & 6.94\\
\hline
Practicality & 8.27 & 8.24 & 8.3\\
\hline
Time & 7.47 & 7.52 & 7.45\\
\hline
Safety & 6.2 & 5.99 & 6.44\\
\hline
\end{tabular} 
\caption{Average priorities by gender} 
\label{tab:prios-genres} 
\end{table} 
\end{minipage}
\end{multicols}

\begin{figure}[ht]
    \centering
    \begin{subfigure}{0.3\textwidth}
        \includegraphics[scale=0.3]{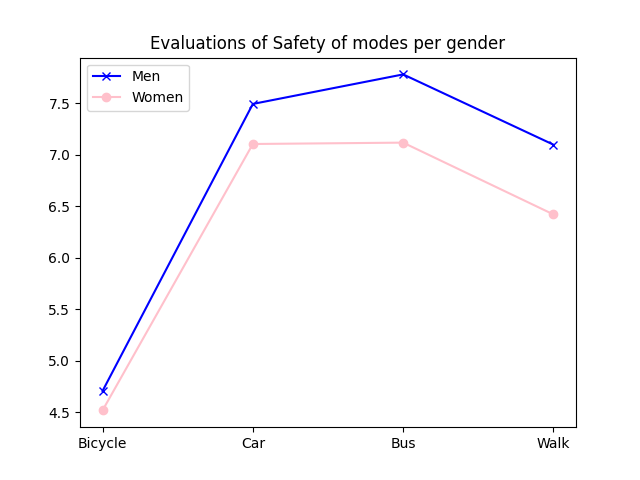}
        \caption{Safety}
    \end{subfigure}
    \begin{subfigure}{0.3\textwidth}
        \includegraphics[scale=0.3]{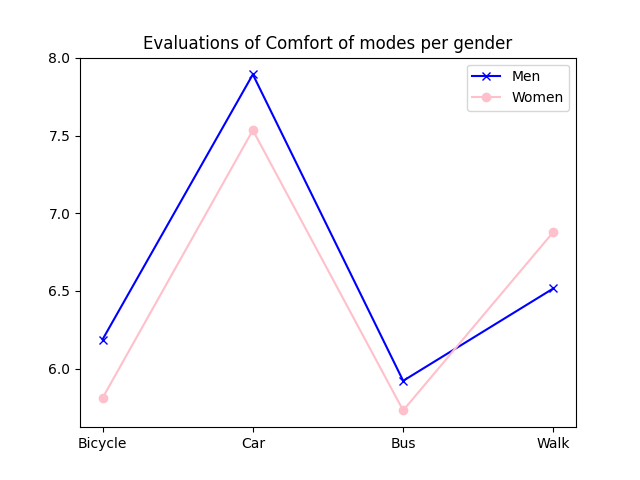}
        \caption{Comfort}
    \end{subfigure}
    \begin{subfigure}{0.3\textwidth}
        \includegraphics[scale=0.3]{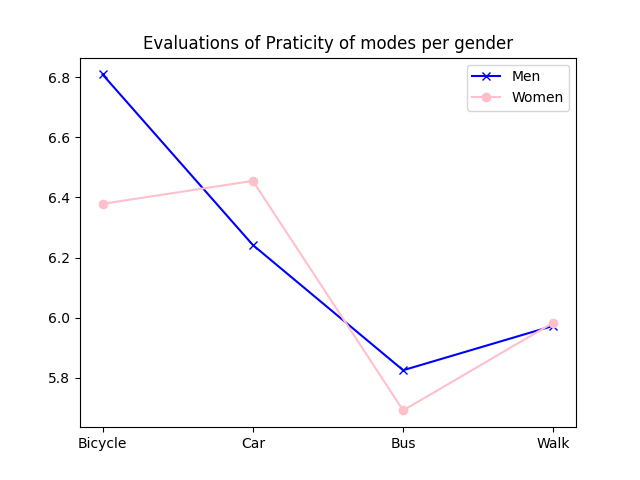}
        \caption{Practicality}
    \end{subfigure}
    \caption{Gendered differences in evaluation of mobility on 3 criteria}
    \label{fig:gender-evals}
\end{figure}

\paragraph{Evaluations.}
To further analyse these results, we show in Table~\ref{tab:tab-hf} the comparison of male vs female evaluations of the 4 modes on the 6 criteria. We can notice that the evaluation of ecology is the most similar between genders. On the contrary, there are sensible differences on the evaluations of Practicality, Safety, and Comfort. Figure~\ref{fig:gender-evals} displays the evaluation differences for these 3 criteria. The differences for safety and comfort match the respective differences in priority for these criteria. We can observe that women rate the safety and comfort of all modes lower, except for walking that they find more comfortable than men. The difference in practicality is more surprising since the priority was the same; it might suggest that due to additional constraints (children, family shopping, etc), the car is really more convenient for women than bicycle or public transport. Indeed, they rate the practicality of bicycle and bus significantly lower than men, and the practicality of car significantly higher, while that of walking is similar.

\begin{table}[ht] 
\centering 
\begin{tabular}{|c|c|c|c|c|c|c|}
\hline 
 & Ecology & Comfort & Finance & Practicality & Time & Safety \\ 
\hline\hline 
Bicycle & (9.27, 9.15) & (6.19, 5.81) & (7.92, 7.5) & (6.81, 6.38) & (6.74, 6.43) & (4.71, 4.52) \\
\hline
Car & (1.85, 1.8) & (7.89, 7.53) & (2.56, 2.8) & (6.24, 6.46) & (6.8, 6.75) & (7.5, 7.11) \\ 
\hline
Bus & (7.48, 7.36) & (5.92, 5.73) & (7.12, 6.6) & (5.82, 5.69) & (5.57, 5.53) & (7.78, 7.12) \\
\hline
Walk & (9.82, 9.83) & (6.52, 6.88) & (9.79, 9.7) & (5.97, 5.98) & (2.69, 3.27) & (7.1, 6.42) \\
\hline
\end{tabular} 
\caption{Gendered evaluations of modes on criteria} 
\label{tab:tab-hf} 
\end{table} 

\begin{figure}[ht]
    \centering
    \includegraphics[scale=0.4]{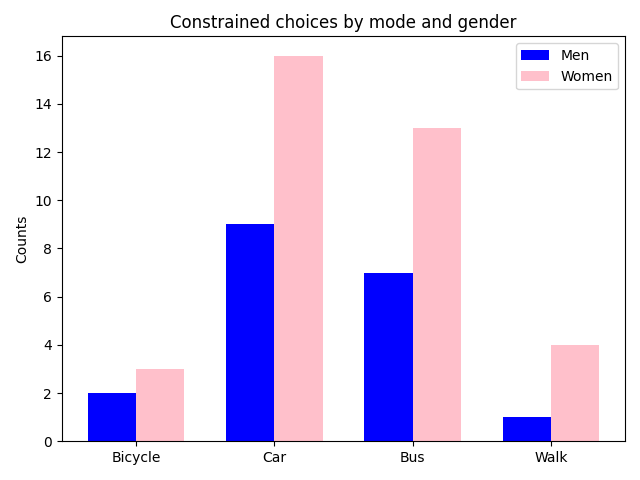}
    \caption{Comparing constrained choices by man vs women}
    \label{fig:gender-constraints}
\end{figure}

\paragraph{Constrained choices.}
Finally, we have compared the number of constrained choices. Concretely, we consider the mobility choice to be constrained, if the rationally best mode, as predicted by the marking algorithm based on declared priorities and evaluations, is listed as unavailable by the responder. For instance, based on a responder's priorities and evaluations, we could deduce that their preferred mode would be public transport, but they reported it as unavailable, maybe because their residence address is not served by any bus line. Among the 650 answers, we have found 59 whose choice was constrained that way, of which only 19 men (5.74 \% of men's answers) and 36 women (11.96\% of women's answers), suggesting that men have less constrained choices than women. Figure~\ref{fig:gender-constraints} reports which modes are selected in this case, by men vs women.

\enlargethispage{20pt}

\newpage
\subsection{Differences per usual mode}

In this paragraph we compare perceptions of mobility depending on the responders' usual commuting mode.

\begin{figure}[ht]
    \centering
    \begin{subfigure}{0.47\textwidth}
        \centering
        \includegraphics[scale=0.52]{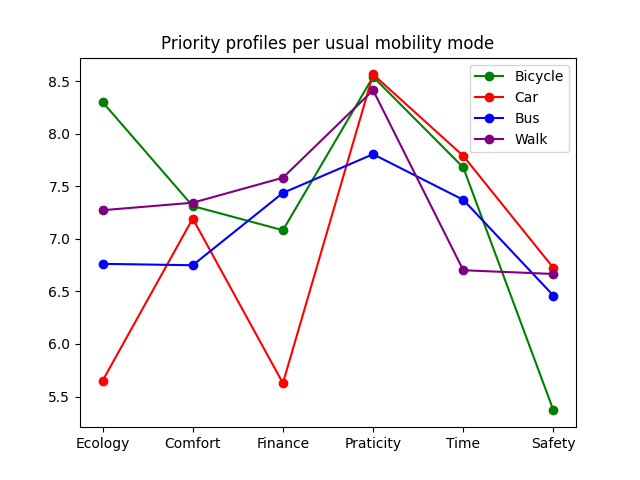}
        \caption{Priority profiles of users of the different modes}
        \label{fig:profiles-prio}
    \end{subfigure}
    \begin{subfigure}{0.47\textwidth}
        \centering
        \includegraphics[scale=0.52]{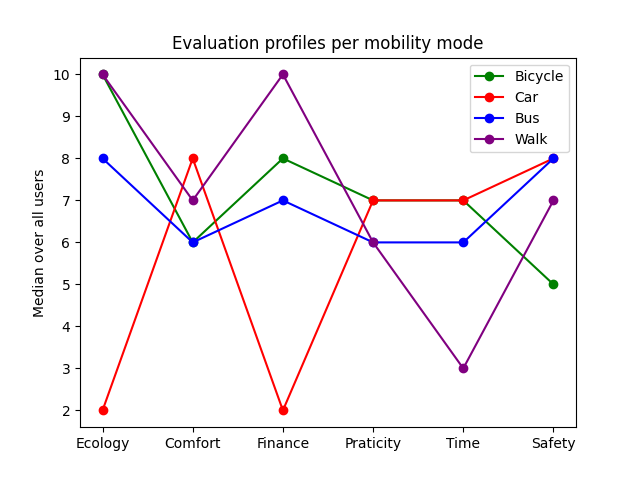}
        \caption{Evaluation profiles of the different modes}
        \label{fig:profiles-eval}
    \end{subfigure}
    \caption{Symmetry in priorities and evaluations per usual mode}
\end{figure}

\vspace*{-20pt}
\paragraph{Priorities.}
Figure~\ref{fig:profiles-prio} shows the difference in average priorities on the 6 criteria, over the populations of habitual users of the 4 different modes. Habitual users are those who declared that mode as their usual commute mode in our questionnaire: over 650 answers, 204 use their bicycle, 134 use their car, 228 use the bus, and 84 walk. The figure shows that some criteria seem to be similarly important for all responders, especially practicality (slightly lower priority for bus users), but also comfort and time (slightly lower for walkers), and therefore less predictive of the differences in choices. On the contrary, we can observe very variable priority profiles over some other criteria. In particular, the priority for ecology ranges from very high over bicycle users, to very low for car users, through average priorities for walkers and bus users. Similarly, car drivers have a significantly lower priority for the price of their mode; walkers have a lower priority for time; and cyclists have a significantly lower priority for safety. We can also notice that the priorities for ecology are quite widespread over the range of values, it is the criterion with the most differences in priorities. On the contrary for finance and safety, only one category diverges from the global agreement. Making a parallel with evaluations of modes on these criteria allow to suggest an explanation.

\paragraph{Evaluations.} Figure~\ref{fig:profiles-eval} shows the median evaluations of the 4 modes on the 6 criteria, computed over all answers. It shows strong similarities between the strengths and weaknesses of a mode over the criteria, and the priority profiles of users of this mode. Said differently, users of a mode will both have a higher priority for the criteria where this mode is globally evaluated as very good and a lower priority for the criteria where this mode is evaluated as bad. We can observe that bicycle is evaluated as highly ecological, and cyclists also declare a much higher priority for ecology. Similarly, car is evaluated very poorly on the financial aspect (it is considered the most expensive, by very far), and car users concomitantly declare a much lower priority for the financial criterion. Besides, this declared priority seems to be in contradiction with the population's concerns for the cost of life, as suggested by polls or by the massive strikes against new tolls or increases in petrol prices. Cyclists use the mode that is considered the less safe by responders, and concomitantly declare a much lower priority for their safety. Yet, safety is one of humans' basic needs, and found to be a strong factor in modal choice \cite{noland1995perceived}. On the other hand, criteria where there is little differences in priority profiles also display low differences in evaluations: all modes are considered similarly comfortable and practical; the bus is evaluated as slightly less comfortable and practical, which coincides with a slightly slower priority of its users for these 2 criteria. Finally, only walking is considered significantly slower, in symmetry with a lower priority for time in walkers. 

\begin{figure}[ht]
    \centering
    \includegraphics[scale=0.5]{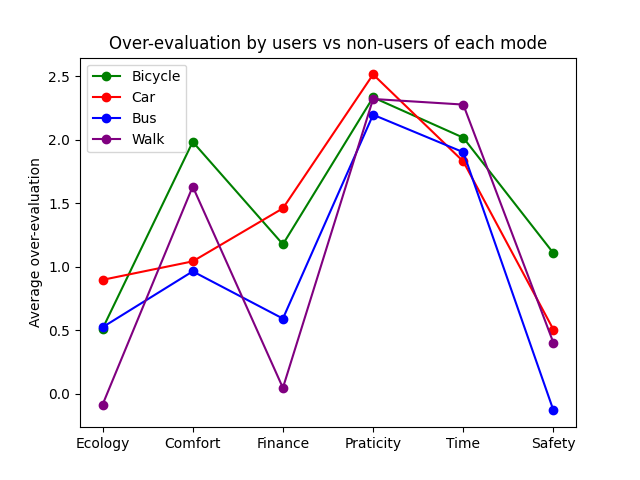}
    \caption{Evaluation difference for each mode between users and non-users}
    \label{fig:overeval}
\end{figure}

\paragraph{Users vs non-users.} 
The figures above show how the general average priorities and median evaluations of the modes on criteria, are symmetric. But even further, users of a mode not only choose a mode whose median evaluation is better on the criteria that are important to them, or whose weak points are not important to them. They also make a different evaluation of their mode compared to non-users. Figure \ref{fig:overeval} illustrates the evaluation difference between users and non-users of each mode, computed as the average evaluation over all habitual users of this mode, minus the average evaluation of non-users (users of the other 3 modes). A positive difference means an over-evaluation by users or an under-evaluation by non-users, while a negative difference indicates an under-evaluation by users compared to non-users. 

We can observe that users of a mode generally over-evaluate it on all criteria compared to non-users. There are 2 notable exceptions, where walkers tend to very slightly underestimate the ecology of walking compared to other users, and bus users slightly under-estimate the safety of buses compared to non-users. There is no deviation in the evaluation of the price of walking, since everybody considers that it is free. Two criteria are equally over-estimated by users of all 4 modes (by over 2 points in average): practicality and time, suggesting that all answers evaluate the mode they use as faster and more practical than others, or choose the mode that they think is faster and more practical. It is interesting to note that these 2 criteria are also the ones reported as being the most important to the participants when choosing a mode. It suggests that the priority of a criterion is linked to its over-estimation by its users.

Other criteria show more differences, with cyclists over-estimating (or non-cyclists under-estimating) more than others their level of comfort and safety ; and car drivers over-estimating how cheap their mode is, and how ecological it is. This partly compensates the lower evaluation of car on these 2 criteria, along with their relatively low priority for these 2 criteria. Said differently, participants who use the car preserve a good score for car by both over-estimating the value and lowering the priority of weak criteria, where car is globally less good than other modes; cyclists preserve a good score for bicycle by neglecting the priority of safety or over-estimating its value. Similar adaptations can be observed on the other modes, in line with our model of the halo bias \cite{isaga-bias}. To conclude, priorities and evaluations of all modes differ depending on each individual's usual mobility mode; these differences are oriented in such a way that priority profiles of users reflect the respective strengths and weaknesses of their usual mode. In the following we suggest some explanations to this symmetry between priorities and evaluations.

\subsection{Explaining mobility decisions}

There are several ways to explain such evaluations and priorities differences between users of different modes.

\paragraph{Context.} First, not all responders to our questionnaire live in the same area, so different evaluations can be explained by different contexts. Not all towns have the same quality of bus network, equal infrastructures to enhance cycling safety, weather conditions, or similar legislation in place about speed limits, which results in different distributions of the population over mobility modes available. For example, towns such as Grenoble or Strasbourg have significantly more cyclists than the national average, due to a long history of urban policies in favour of cycling. However, some criteria do not depend on the regional context, such as ecology (walking or cycling are always more ecological than driving), or have a limited regional dependency, such as price (car insurance and petrol price are expensive in all areas) and time (walking is slower than the other modes beyond very short distances). Therefore, differences on the evaluation of these criteria cannot be explained by the context, which suggests that other explanations can be found.

\paragraph{Knowledge by usage.} Second, people who do not use a mode are less likely to have an accurate evaluation of its qualities, and more likely to over-evaluate its risks or drawbacks. Said differently, someone needs to try a mobility mode before they can have an informed opinion about it. For instance, people will tend to overestimate walking time, in particular when they are less physically active \cite{dewulf2012correspondence}; such overestimation might deter them from walking \cite{ralph2020really}; some towns are thus investigating the use of signs for pedestrians that indicate the walking time to different monuments, in order to encourage walking \cite{borowik2019evaluation}. Similarly, our results show that non walkers under-estimate the speed criteria of walking, or that non cyclists find the bicycle to be less comfortable and more dangerous than cyclists do (in line with fear of the unknown as discussed in \cite{bcg2020}). In the other direction, bus users find the bus less safe than non users, suggesting that they might be more aware of the potential risks since they have witnessed them directly. But this does not explain why evaluations and priorities deviations are aligned.

\paragraph{Rational selection.} Third, it seems obvious that people who have a better perception of a mode are more likely to use it: people who find that cycling is very comfortable are more likely to use a bike than people who think it is very uncomfortable, all the more if comfort is important to them. It is therefore only natural that users have a globally better perception of their mode than of the other modes. However, some differences are hard to justify this way when there is some sort of objective evaluation (cycling or walking is objectively very ecological; driving is objectively quite expensive), or when their amplitude is too wide.

Besides, this deviation is generally accompanied by a similar deviation from the average priority: car drivers have a lower priority for ecology and finance, where their mode performs poorly compared to others; cyclists report a significantly lower priority for safety, where bicycle is evaluated as more dangerous; or walker report a lower priority for time, while using the slowest mode. Assuming objective evaluations of a mode on criteria, it seems logical that each mode will attract users whose priorities align with the evaluations, \ie users who have a higher priority for their qualities, and lower priorities for their weak points. For instance, walking or cycling being the least expensive modes and most ecological are likely to attract users very concerned with price and/or ecology; inversely, walking is very slow so it is not likely to attract users interested in shortening their trip time, while car being very expensive and less ecological will rather attract users not concerned with price or ecology. However, some priorities reported do no seem consistent: it is not likely that all drivers actually do not care about price, or that all cyclists do not care about their safety, suggesting that they could modify these priorities \emph{a posteriori} to support their mobility choice.

\paragraph{Biases.}
Finally, some differences could also be explained by cognitive biases. The \textbf{choice-supportive bias} or post-purchase rationalisation is well-known in marketing, and represents the tendency to retroactively amplify the positive attributes of the selected option, or the negative attributes of the ignored options. The goal is to minimise regrets after a choice. Applied to our context, this bias would push people to give better marks to their mode and worse marks to the others, in order to feel better about their choice, which is consistent with our observations. The \textbf{halo bias} consists in selectively focusing on some aspects to preserve a first impression. Concretely here, people could adjust their priorities to focus on strong aspects and neglect weak aspects of their chosen mode, in order to preserve their satisfaction and avoid a costly modal change. This is also consistent with our observations so far. In the sequel we explore how our observations can support the existence of these cognitive biases in mobility decisions. Please note however that these can only be presented as hypotheses, as other explanations could lead to the same observations, as discussed above. Cognitive biases are proposed here as one potential explanation, in order to support the necessity to take them into account when designing urban policies to encourage soft mobility.

\section{Cognitive biases in mobility decisions} \label{sec:biases}

\subsection{Post-purchase rationalisation bias}

\begin{figure}[ht]
    \centering
    \begin{subfigure}{0.24\textwidth}
        \centering
        \includegraphics[scale=0.3]{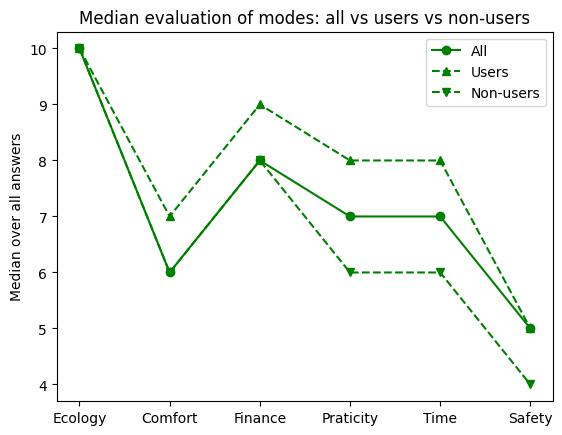}
        \caption{Bicycle}
        \label{fig:wisdom-bike}
    \end{subfigure}
    \begin{subfigure}{0.24\textwidth}
        \centering
        \includegraphics[scale=0.3]{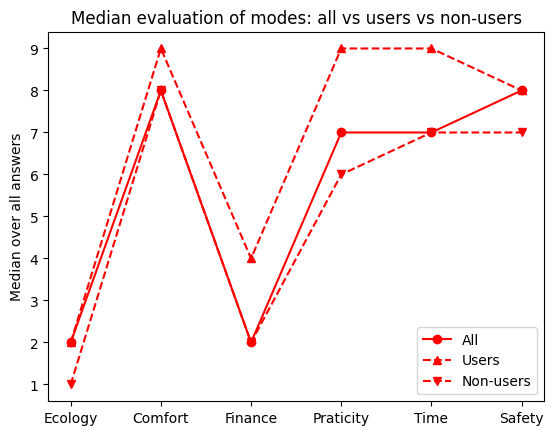}
        \caption{Car}
        \label{fig:wisdom-car}
    \end{subfigure}
    \begin{subfigure}{0.24\textwidth}
        \centering
        \includegraphics[scale=0.3]{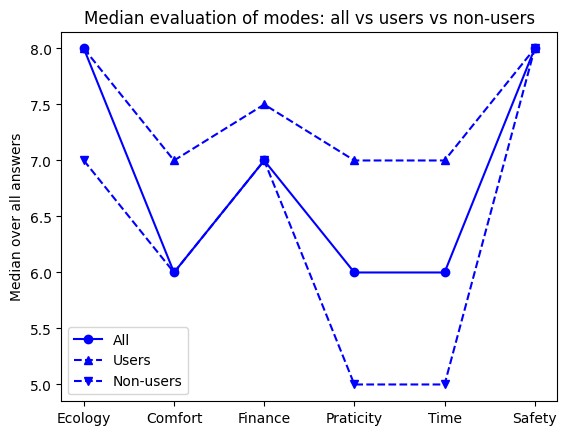}
        \caption{Bus}
        \label{fig:wisdom-bus}
    \end{subfigure}    
    \begin{subfigure}{0.24\textwidth}
        \centering
        \includegraphics[scale=0.3]{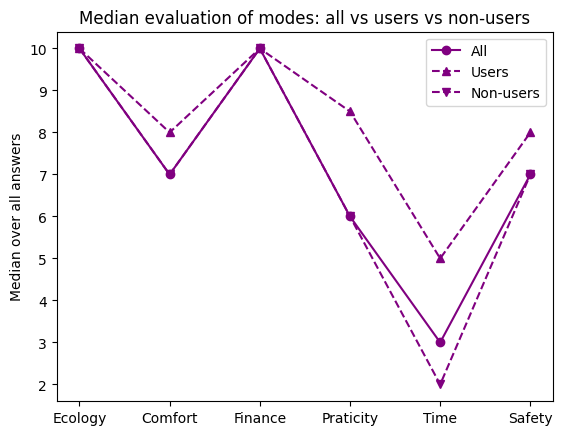}
        \caption{Walk}
        \label{fig:wisdom-walk}
    \end{subfigure}
    \caption{Evaluation medians for all, users, non-users}\label{fig:wisdom}
\end{figure}

\paragraph{Wisdom of the crowd.}
From the observations above, it is difficult to deduce the 'objective' value of each mode on each criteria, or said differently it is hard to know if users over-evaluate their mode, if non-users under-evaluate it, or if they are all right and the different evaluations come from differences in their context. Wisdom of the crowd \cite{galton1949vox,surowiecki2005wisdom} is the idea that when evaluating a continuous value, the collective evaluations of many individuals could be considered as a probability distribution, with the 'real' value close to the median. Applied here, we could try to 'crowdsource' the objective evaluation of each mode on each criterion, in order to counter the subjective deviations noticed between users and non-users. Figure \ref{fig:wisdom} illustrates how users and non-users evaluate each mode on all criteria, compared to the median evaluation. This figure makes evaluation variations even more visible, in particular the differences between users and non-users in evaluating practicality and time of bus trips. Besides, we can observe both over-evaluations from users, and under-evaluations from non-users, for most criteria. This tends to confirm the influence of the post-purchase rationalisation bias, where people over-evaluate good aspects of a choice and under-evaluate its bad aspects.

\paragraph{Rationality of decisions.}
We then decided to check if the mobility choices declared by the responders to our survey were rational, in the sense of our multi-criteria decision algorithm explained above (Section~\ref{sec:ratio}). Concretely, we considered a choice as rational if it receives the best score among all the available modes for this individual. In a first method, we compared the scores computed with the individual's self evaluations of modes on each criteria. However, as shown above, these evaluations could be biased so as to rationalise the decision \emph{a posteriori}. Therefore in a second method, we compared the marks obtained with the individual priorities but the crowd wisdom, \ie with the median evaluation of modes on criteria over the entire population of responders, in order to cancel evaluation biases.

\begin{figure}[ht]
    \centering
    \includegraphics[scale=0.4]{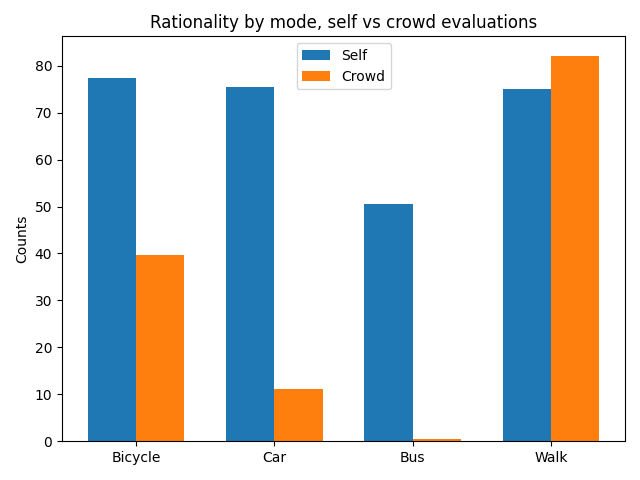}
    \caption{Percentage of rational choices, per mode, with self evaluations vs crowdsourced evaluations}
    \label{fig:bar-ratio}
\end{figure}

Figure~\ref{fig:bar-ratio} reports the percentage of rational decisions with both methods. First, we can observe that most decisions are rational with self-evaluations: about 80\% choices are rational for users of bicycle, car, and walk. However, the choice to use the bus is only rational for half of its users. Second, we can observe that significantly less decisions are rational when using crowd evaluations. Indeed, this cancels the post-choice rationalisation and therefore reveals irrational choices that were hidden (or justified) by this bias. Another explanation would be that the median mark is under-estimating the real value of a mode: indeed, according to the crowd, no decision to take the bus is ever rational. On the contrary, crowd evaluations make walking even more rational than self evaluations. In conclusion, individual evaluations of modes on criteria make the individual decisions more often rational than the crowd evaluations, which suggests that individuals could adapt their evaluations to rationalise their choices.

\paragraph{Remark.}
It is interesting to notice that to evaluate the rationality of the responder's choice, we considered only the options that they declared as available to them. However, these available options are also subjective and could be biased to rationalise the user's choice. For instance, someone can consider that public transport is not available because it is 'too far', or over-evaluate the walking time to reject this option, or feel that cycling is not available due to their fitness level which could change. Future work would be needed to investigate these other perception biases.

\subsection{Halo bias}

Our rational algorithm only accounts for about 80\% of the mobility choices of our responders. In order to capture more of the reported modal choices, we investigated another bias: the halo bias, which consists in focusing on the good aspects of one's choice, and neglecting drawbacks. Concretely, we assume that individuals tend to adapt not only their evaluations (under the effect of the choice-supportive bias), but also their priorities (under the effect of the halo bias), to justify their usual modal choice.

\begin{figure}[ht]
    \centering
    \includegraphics[scale=0.4]{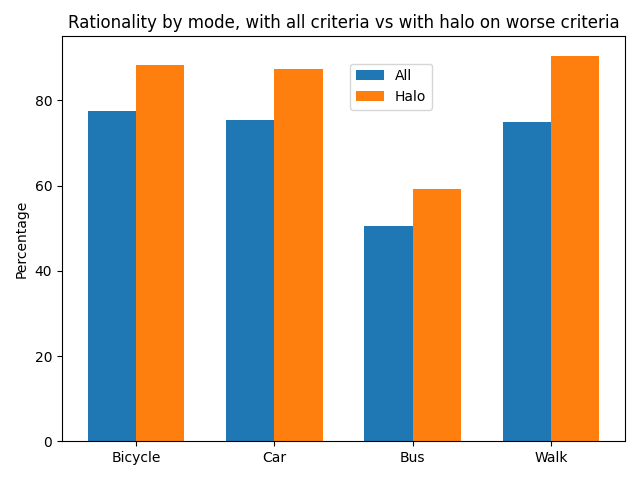}
    \caption{Percentage of rational choices, per mode, with all criteria vs with halo on worse criteria}
    \label{fig:halo-bars}
\end{figure}

Figure~\ref{fig:halo-bars} shows the percentage of decisions that are considered as rational based on our multi-criteria algorithm, using all 6 criteria (blue bars, same as Figure~\ref{fig:bar-ratio} above). Besides, it also shows the percentage of decisions that are rational if applying the same algorithm, but neglecting some criteria, \ie applying the halo bias. Concretely, for each response, we re-computed the average mark with a selected set of criteria, ignoring those where the declared mode had the worst evaluation (thus setting the priority to 0 for these criteria), in line with our model \cite{isaga-bias}. We can observe that the rationality of the bus still remains quite low. Indeed, bus users tend to have a more balanced evaluation, where the bus does not stand out on any criterion, and symmetrically no priority stands out either. As a result, no single criterion can be ignored to improve the global mark much. But globally, the halo bias allows to capture more decisions than just the rational multi-criteria decision algorithm. This supports our hypothesis that the halo bias can explain some decisions that initially looked irrational. In future work, we could also investigate different implementations of the halo bias, such as focusing only on the best rated criteria, or on a limited set of the most important criteria.

\begin{table}[ht]
    \centering
    \begin{tabular}{|c|c|c|c|c|c|c|}
    \hline
    Mode & Ecology & Comfort & Finance & Practicality & Time & Safety \\
    \hline \hline 
    Bicycle & 0 & 6 & 1 & 2 & 1 & \textbf{19}\\ \hline
    Car & \textbf{15} & 0 & 9 & 0 & 0 & 0\\ \hline
    Bus & 3 & 8 & 9 & 5 & 4 & 2\\ \hline
    Walking & 0 & 0 & 1 & 0 & \textbf{12} & 3 \\ \hline
    \end{tabular}
    \caption{Number of decisions made rational by ignoring each criterion among users of each mode}
    \label{tab:halos}
\end{table}

\enlargethispage{20pt}

Further, Table \ref{tab:halos} shows which criteria are ignored per mode, with the count. Unsurprisingly, different criteria stand out for each mode: cyclists can put a halo on safety; car drivers could ignore ecology and finance; bus users would lower their attention to comfort and price; and pedestrians mostly need to lower their focus on time. Halo bias, as we model it, is complementary to the post-purchase rationalisation bias: individuals tend to both over-evaluate their mode on some weak criteria, and completely ignore the worst-rated criteria to justify their mode. However, this is still not sufficient to capture all decisions.

\subsection{Reactance bias}

In order to visualise potential reactance biases, we focused on how users of specific modes evaluate the other modes, instead of just taking the average over all non-users. The goal behind this is to reveal particular 'dislikes' between some modes, that could be due to official messages or policies that seem to favour one or the other, or seem to push users to change mode. In particular, cyclists are often perceived very negatively by car drivers \cite{delbosc2019dehumanization}, and vice-versa \cite{paschalidis2016put}. New mobility modes such as e-scooters are also perceived negatively by pedestrians \cite{james2019pedestrians}, but unfortunately these new modes were not part of our survey.

\begin{figure}[ht]
    \centering
    \includegraphics[scale=0.5]{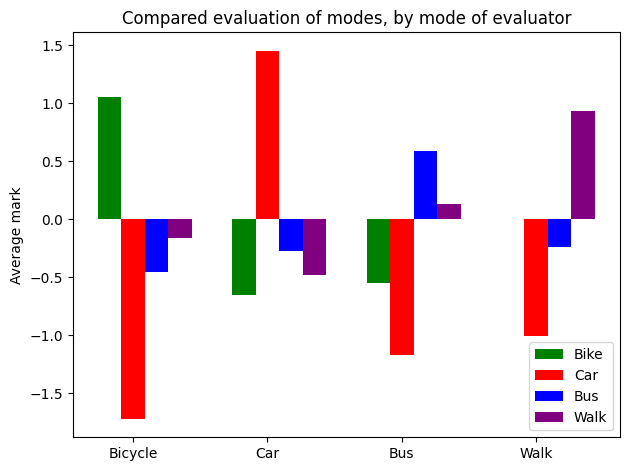}
    \caption{Deviations in evaluation, by users of each specific mode}
    \label{fig:reactance}
\end{figure}

Figure~\ref{fig:reactance} shows how users of each mode (represented by the colored bars) over- or under-evaluate each mode. On the left, we can see how bicycle is evaluated by the users of the 4 modes. With no surprise, cyclists do over-evaluate their own mode, while the users of the other modes under-evaluate it. But what is new here is that we can see that the car drivers do under-evaluate bicycle far more than walkers or bus users. Yet, there is no objective reasons while drivers would precisely have such lower perceptions than users of public transports or pedestrians. Alternatively, we suggest that this expresses some sort of reactance bias where drivers feel questioned or blamed by the public discourse that they should use their bicycle. By reactance, they under-rate it even more in reaction, to assert their freedom of choice. This is in line with studies showing aggressive relationships of car drivers to cyclists \cite{delbosc2019dehumanization} perceived as obstacles, or not considered to be legitimate on the road \cite{oldmeadow2019driver}. The other modes (public transport and walking) are less bothersome to drivers, as they are not conflicting for the same space (driving lanes), which is reflected by lower deviations in their evaluation; they have also been around for longer than newer modes such as rental bikes or e-scooters. Pedestrians also have potentially negative interactions with cyclists, yet they do not under-evaluate bike as a mobility mode as much as motorists; indeed they cannot feel blamed by official communication campaigns since they already use a soft mobility mode. Symmetrically, cyclists are those who under-evaluate car driving the most, although this is less impressive than drivers' under-estimation of cycling.

\enlargethispage{15pt}
We can observe that car drivers tend to be the less objective judges: they over-estimate their own mode and under-estimate the other modes far more than others. This could be explained by a very different priority profile, where car users tend to neglect ecology and money, which are quite important to the average responders in our sample. Inversely, pedestrians seem to be more objective, slightly under-evaluating bicycle and car, and even over-estimating slightly the bus; this is consistent with the fact that they are not blamed for their mobility choices, unlike car drivers accused of contributing to global warming, and cyclists or users of emerging mobility modes, accused of not respecting safety rules \cite{johnson2013cyclists,fraboni2018red}. 

\begin{figure}[ht]
    \centering
    \begin{subfigure}{0.45\textwidth}
        \centering
        \includegraphics[scale=0.4]{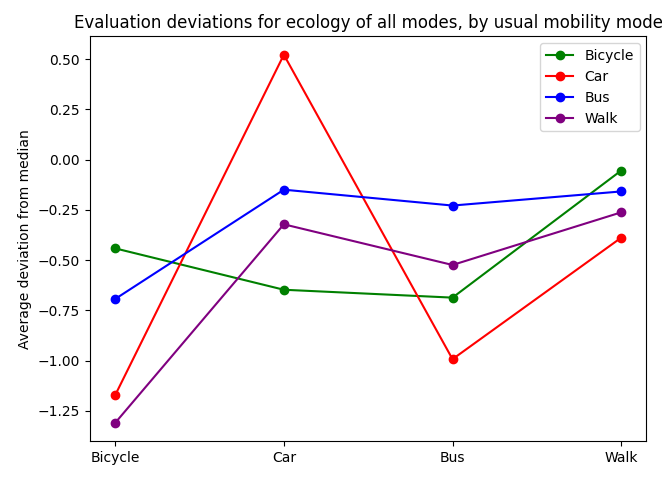}
        \caption{Evaluation deviations for ecology of modes}\label{fig:dev-eco}
    \end{subfigure}
    \begin{subfigure}{0.45\textwidth}
        \centering
        \includegraphics[scale=0.4]{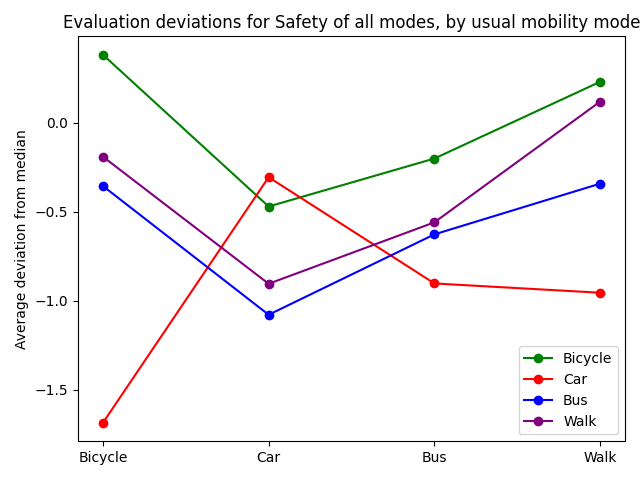}
        \caption{Evaluation deviations for safety of modes}\label{fig:dev-safe}
    \end{subfigure}
    \caption{Mean evaluation deviation by usual mobility mode}
    \label{fig:deviations}
\end{figure}

\paragraph{Ecology.} Since most official communication is about more ecological or sustainable mobility choices, we further focused on the perceptions of the different modes on this specific criterion. Figure~\ref{fig:dev-eco} shows the mean deviations between the evaluation of ecology of each mode, by users of each mode. For instance, the red line shows how car drivers evaluate the ecology of each mode. We can notice that they over-evaluate the ecology of cars, while they under-evaluate that of cycling or public transport, and slightly that of walking. Interestingly, walkers also under-evaluate the ecology of cycling, which might reveal a reactance bias, since those 2 categories of users are those that have the most conflicts with cyclists who share their mobility space. Apart from this, users of bicycle, bus and walk have quite similar evaluations of ecology of all modes.

\paragraph{Safety.} We also decided to focus on safety, another topic of official communications to encourage safe sharing of the urban space. Figure~\ref{fig:dev-safe} shows the results. We can observe a very different evaluation of safety by car drivers with respect to all other users. They particularly under-estimate the safety of bicycles and pedestrians. Cars and bus are considered the safest modes, in line with road deaths statistics. Bus users are more concerned with the risks of aggression than collision, and their evaluation of bus safety is significantly higher among men (mean 7.78) than among women (mean 7.12), as discussed above (Section~\ref{sec:gender}).

\begin{comment}
\begin{figure}[ht]
    \centering
    \includegraphics[scale=0.5]{imgs/compar-safety.png}
    \caption{Comparing mean safety evaluations by usual mobility}
    \label{fig:safety}
\end{figure}
\end{comment}

\paragraph{Cost.} Some responders also reported that walking was expensive (5 people rated it at 0 for financial accessibility, 19 in total rated it below 7), and many reported that cycling was expensive, while the costs are very limited. This could also be interpreted as a reactance bias where people mark bicycle severely in reaction to insistent promotion campaigns. Future work will be dedicated to further analyse answers, in particular the text comments, to study the influence of reactance.

\section{Discussion} \label{sec:discuss}

\subsection{Limitations of the survey}

\paragraph{Formulation.}
There are several limits in our survey. First, some questions were ambiguous or understood differently by different responders. For instance, we did not exploit the daily distance travelled or number of weekly trips due to too many abnormal answers, with values up to 2000km home-work distance, or 40 trips per week. The inaccessibility of mobility modes can also be understood differently, either as a complete lack of availability vs an unsuitability for one's personal constraints: it is different to not be able to walk at all, vs considering that one's work is too far to walk. Second, our attention to not collect personal or identifying data does limit our knowledge of the particular context of each individual: age and town would have provided useful information to further analyse the answers, and to attribute decisions to contextual factors where relevant. Third, we only studied four mobility modes, which reduces the reach of our study. We do not account for emerging mobility modes such as e-scooters. Further, we did not specify whether 'bicycle' represented a muscular bike or an electric-assisted one. Similarly, the 'car' mode does gather all sorts of vehicles, electric or not, which have different attributes. This leads to different evaluations by respondents depending on their interpretation of the questions. 

\paragraph{Sample.}
Finally, our sample was biased, as discussed above. In other work, we have corrected this bias to initialise a virtual population representative of the national French population, from values obtained in the survey. But in the context of the present study, which is targeted at identifying biases, we consider that the sample serves our purpose. Indeed, biases can affect any user, independently of their mobility mode.

\paragraph{Decision factors.}
Our survey relies on 6 decision factors that were previously identified. In the literature, other surveys use similar factors. For instance \cite{weyer2023modeling} clustered users based on almost the same 6 factors,  with the exception of the 'reliability' factor replacing our 'practicality' factor. The decision dimensions retained in such models are all coarse-grained approximations of more precise determinants that depend on each travel mode: congestion or availability of parking for car driving, reliability or connectivity of public transport \cite{tyrinopoulos2013factors}, or altitude profile when walking or cycling. We suggest that our idea of practicality, which could also be named 'fluidity', captures specific aspects of the different modes that impact the user's cognitive load: continuity of cycling lanes (no thinking about itinerary), continuity and reliability of bus lines (no connections, no risks of failure), continuity of car trips (no stop and go in traffic jams). Walking looks like the 'fluidest' mode, but weighed down by its slowness. Besides, other factors are not considered in our survey nor in the literature. In particular, the health impact is becoming more important after the pandemics, encouraging to choose active mobility modes as a way to get physical exercise \cite{scorrano2021active}. Future work could survey the relative importance of these other criteria, but the current survey was sufficient to reveal the influence of cognitive biases.

\subsection{Cognitive biases as potential explanations}

Another important point that we want to insist on is that the cognitive biases identified here are presented only as hypotheses. We suggest them as plausible explanations of the observed 'car stickiness', or inertia in mobility despite new policies and infrastructures aimed at favouring soft mobility. It is impossible to prove for a given individual what really caused a decision, all the more without details about their personal context, since we did not gather any personal information. We can only conclude that biases that capture a large number of decisions are plausible explanations for similar variations observed in many individuals, and should thus be taken into account by deciders. This is in line with other works such as Fouillard \etal \cite{fouillard2021catching}, who propose a logical model to diagnose cognitive biases in erroneous decision making, applied to plane accidents. Instead of asserting that one or the other bias is the only explanation to the accident, their model allows to infer all possible biases that could explain the observed scenario. Our study here fits this approach: we describe cognitive biases that explain the observed behaviours, without asserting that they are necessary or sufficient explanations. 

Besides, several different explanations can interact together, for instance a choice can start as constrained (such as using a bicycle before owning a driving license) before being rationalised \emph{a posteriori} to preserve some satisfaction (\eg insisting on the ecological value of this choice). A choice can also be rational initially (such as driving to work when the distance is too long and no practical public transport is available), and later protected when the urban environment evolves (\eg new bus lines), by adopting biases (\eg over-estimating the risks of taking the bus, such as strikes or delays), to avoid a costly modal switch (\eg giving up to car and take the bus). A choice can come from a routine (\eg drive to work as a cultural reflex), but be reinforced by modifications of priorities that justify it (\eg insist on the importance of comfort or autonomy).

Finally, other biases not studied here can also play a role, in particular in reaction to communication campaigns. Confirmation bias, anchoring, or reactance can modify people's beliefs and prevent them from accepting official messages, as seen in climate change denial, or lack of respect of road safety rules. Advertisement campaigns play a role in giving a good image of the car, symbol of autonomy, adventure, power, or social success. Rebound effect should not be neglected either: measures that initially seem favourable can then turn around: building larger roads to decrease traffic jams will eventually attract more cars; telecommuting will reduce the number of trips, but also push people to live further away from their workplace and increase the distances traveled; making cars more energy-efficient will reduce the cost per kilometre and eventually lead to more trips. But our model as it is already provides useful insight about mobility decisions, as explored in the next paragraph.

\subsection{Applying our model}

One point of interest with models of modal choice is the ability to deduce modal transfer to new modalities, or to existing ones when their properties change. For instance, urban planners may want to predict if building an urban cable transport will reduce the number of cars, \ie if the future cable transport users will be former motorists, or rather former users of public transport or pedestrians. Similarly, who would take public transport if they were made free? Would more car drivers switch to cycling of cycling lanes were made more secure? In our survey, we have gathered the perceptions and priorities of users, which allow us to compute the rational choice for them in their current context, but also if the context would change. Concretely, we can introduce a new value for some mode on some criterion, simulating an urban planning policy, and recompute the resulting rational choices of all users. Other work have studied similar scenarios based on such a model: for instance \cite{scorrano2021active} modify values in their model to predict the impact of policies change in various scenarios, such as building more cycling lanes.

\begin{figure}[ht]
    \strut\hspace*{-05pt}
    \begin{subfigure}{0.3\textwidth}
        \includegraphics[scale=0.35]{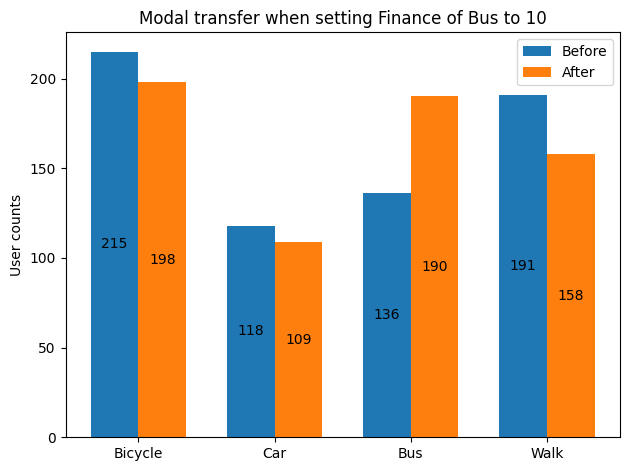}
        \caption{Free public transport}
    \end{subfigure}
    \begin{subfigure}{0.3\textwidth}
        \includegraphics[scale=0.35]{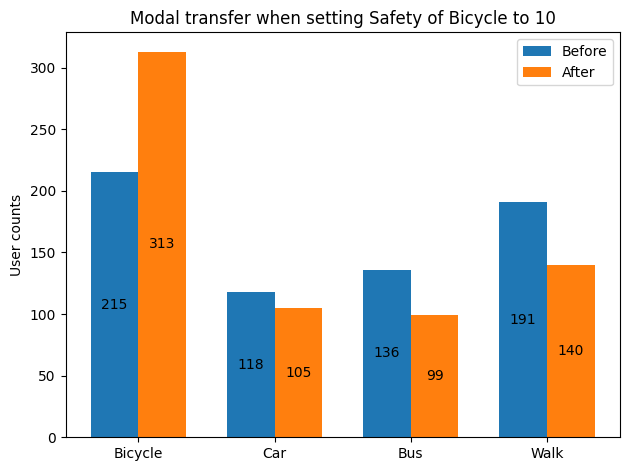}
        \caption{Safe cycling lanes}
    \end{subfigure}
    \begin{subfigure}{0.3\textwidth}
        \includegraphics[scale=0.35]{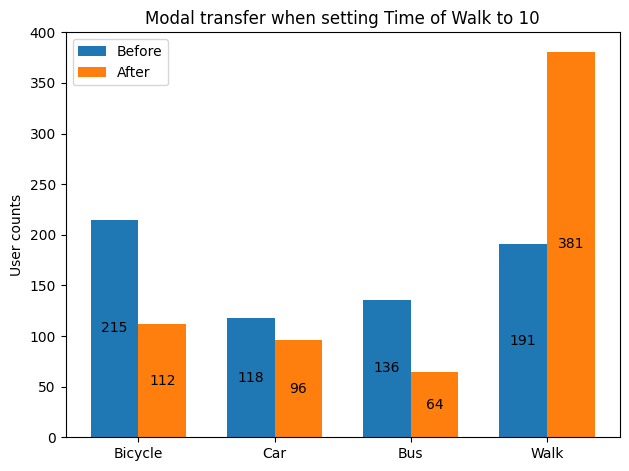}
        \caption{15-minute city}
    \end{subfigure}
    \caption{Predicting modal transfer with various policies}
    \label{fig:policies}
\end{figure}

Figure~\ref{fig:policies} illustrates the predicted modal transfers with three different urban policies. On the left, public transport is made free, which is represented in our model by an evaluation of 10 on the finance criterion. We can see that the bus gains new users, mostly at the expense of walking. In the middle, there are safe cycling lanes, which is modelled as an evaluation of 10 for bicycle on the safety criterion. As a result, there are more cyclists but mostly former pedestrians, and some bus users, but very few former car drivers. Finally on the right, we have modelled the idea of a 15-minute city, \ie a city where everything is in walking distance range, by setting the evaluation of walking on the time criterion to 10. Again, many users switch mode, but mostly cyclists and bus users. Again, car drivers mostly kept to their initial choice. These are only three examples, but it shows that users will choose modes that are similar to their preference profile; if other modes are improved on criteria that they are not interested in, this will not have much impact on their decisions. Thus for instance, car drivers who have lower priorities than the average on finance will not react to free public transport.

Of course this is only a simplification and the prediction cannot be considered fully valid, but it illustrates the power of such a model once correctly calibrated on the real population. It also shows that given their values (or priorities for the criteria), not all modal transfers are equally likely. Said differently, emerging mobility modes are likely to gain users at the expense of walking or bicycle, and sometimes bus, rather than among car drivers, whose values are too different. Besides, this does not even take habits or cognitive biases into account, which might induce more inertia in the modal transfers even though the rational choice might have evolved. In other work, we use a simulator based on this model to illustrate the impact of cognitive biases, by letting the user switch them on or off in the decision making of the virtual population. This way, the user can observe the difference in impact of their urban policies depending on cognitive biases, and realise that they can create an inertia in the adoption of new mobility modes \cite{isaga-bias,isaga-habits}.

In future work, it would be interesting to compare such predictions with real observations in cities that have enforced new policies, for instance free public transport in Dunkerque \cite{briche2017dunkerque} or more recently metropolitan train in Strasbourg \cite{rousseau2023premiers}, in order to validate the model. Further, we would also like to apply clustering algorithms in order to find classes of similar user profiles based on their declared preferences, which would help tailoring policies or communication to their needs and expectations, making them more efficient. Finally, ongoing work is dedicated to the design of a serious game based on this model, where the player will take on the role of a land-use planner who can decide urban policies in order to make mobility more sustainable in a virtual city.

\section{Conclusion} \label{sec:cci}

In this paper, we have described a survey about the perceptions of 4 mobility modes and the preferences of users for 6 modal choice factors. This survey has gathered 650 answers in 2023, that are published as open data \cite{ELLXJF_2024}. In this study, we have used these results to highlight the influence of a number of cognitive biases on mobility decisions: halo bias, choice-supportive bias, and reactance. These cognitive biases are proposed as plausible explanations of the observed behaviour, where the population tends to stick to individual cars despite urban policies aiming at favouring soft mobility. We have discussed how this model and the gathered data can also serve as the basis to simulate how a virtual population reacts to urban policies enacted by a player. Other work has been dedicated to implementing agent-based models of mobility choice, influenced by various biases \cite{isaga-bias}, and informed by this data \cite{conrad2024identifying}. Work is still ongoing to design a simulation-based serious game where the player takes the role of an urban manager faced with planning choices to make their city more sustainable. Future work is also planned to further analyse the data, in particular to deduce clusters of similar users, and to analyse the qualitative answers to open-text questions. This survey provides useful insight into human reasoning in response to urban policies, that can inform agent-based modellers or policy makers.

\section*{Acknowledgements}

This work is part of the ANR project SwITCh funded by the French National Research Agency under number ANR-19-CE22-0003.

\footnotesize
\bibliographystyle{plain}

\end{document}